# Which notes are Vadi-Samvadi in Raga Rageshree?

[1]Soubhik Chakraborty*, [2]Rayalla Ranganayakulu, [3]Shivee Chauhan, [4]Sandeep Singh Solanki and [5]Kartik Mahto

[1]Department of Applied Mathematics, BIT Mesra, Ranchi-835215, India

[2, 3, 4, 5]Department of Electronics and Communication Engineering,

BIT Mesra, Ranchi-835215, India

**\*Soubhik Chakraborty is the corresponding author; email:** soubhikc@yahoo.co.in
Telephone: +919835471223

Email addresses of other authors are as under:-

rayalla.ranga@gmail.com (R. Ranganayakulu)
shivee.chauhan@gmail.com (S. Chauhan)
sssolanki@bitmesra.ac.in (S. S. Solanki)
kartik_mahto@rediifmail.com (K. Mahto)

## Abstract

The notes which play the most important and second most important roles in expressing a raga are called Vadi and Samvadi swars respectively in (North) Indian Classical music. Like Bageshree, Bhairavi, Shankara, Hamir and Kalingra, Rageshree is another controversial raga so far as the choice of Vadi-Samvadi selection is concerned where there are two different opinions. In the present work, a two minute vocal recording of raga Rageshree is subjected to a careful statistical analysis. Our analysis is broken into three phases: first half, middle half and last half. Under a multinomial model set up holding appreciably in the first two phases, only one opinion is found acceptable. In the last phase the distribution seems to be quasi multinomial, characterized by an unstable nature of relative occurrence of pitch of all the notes and although the note whose relative occurrence of pitch suddenly shoots is the Vadi swar selected from our analysis of the first two phases, we take it as an outlier demanding a separate treatment like any other in statistics. Selection of Vadi-Samvadi notes in a quasi-multinomial set up is still an open research problem. An interesting *musical cocktail* is proposed, however, embedding several ideas like melodic property of notes, note combinations and pitch movements between notes, using some weighted combination of *psychological* and *statistical* stability of notes along with watching carefully the sudden shoot of one or more notes whenever there is enough evidence that multinomial model has broken down.

# 1. Introduction

The notes which play the most important and second most important roles in expressing a raga are called Vadi and Samvadi swars respectively in (North) Indian Classical music. In an earlier work [1], a strategy for objectively deciding the Vadi swar was proposed where it was stated that a note in order to be objectively classified as a Vadi Swar should satisfy the following two conditions (see the discussion in sec 3 why this note cannot be the tonic Sa even if it meets the conditions; Sa can, however, be a Samvadi):

(i) it should exhibit a high relative frequency of occurrence of its pitch and

(ii) in addition to (i), the relative frequency of occurrence of pitch should stabilize in the shortest period of time.

The first condition is necessary in that an important note, whether Vadi or not, cannot afford to have low relative frequency. However condition (i) cannot be sufficient because a stay note ("nyas swar") can also have a high relative frequency without being Vadi or Samvadi(e.g. Sudh Re in Yaman[1]). Again, a raga can have several nyas swars and the Vadi may itself be a nyas swar(e.g. raga Durga of Khambaj thaat where Vadi: Sudh Ga; Samvadi: Sudh Ni; nyas swar: Sudh Ni, Sudh Ga, Sudh Ma) or it may not be(e.g. raga Durga of Bilaval thaat where Vadi: Sudh Ma; Samvadi:Sa; nyas swars: Sudh Re, Pa and Sudh Dha). Hence condition (i) can be grossly misleading without (ii). On the other hand although (ii) is a sufficient condition, it is to be assessed in addition to the necessary condition (i) and not independently because it is quite possible that an unimportant note, thanks to the performer, can be accidentally stable but such a note will not have a high relative frequency and should be ruled out of the contest straightaway for being Vadi (the reader will have an evidence of this in this very paper!). In summary, we first select the important notes, (Sa excluded of course; see sec 3), which satisfy (i) and then, among the selected lot, we pick up the one which satisfies (ii). This is the Vadi Swar at least *by performance*. Now, there can be one of the following interesting situations:

1. We already know which swar is Vadi from theory and the experimental results confirm it (e.g. the Pilu performance by the first author analyzed in [1])

2. We already know which swar is Vadi but the experimental results contradict it. Although such a case is yet to be reported, if it happens, it would not be correct to throw away the theory. It is more correct to infer that the theoretical claim is not reflected in the performance. That said, we would then proceed to analyze the recordings of other performers, preferably both vocal and instrumental, and check out how many of them contradict the theory. Only then can some decision be taken.

3. There is a difference of opinion in Vadi-Samvadi selection such as in Bageshree, Bhairavi, Shankara, Hamir, Kalingra and Rageshree (the last one analyzed here) and using our objective analysis, we are either able to resolve the conflict or find both opinions acceptable. Accepting two contrasting opinions in Indian Classical music may be rational, especially if the opinions are *Gharana* (school of music) based. But even then, an opinion that characterizes a specific Gharana ought to be reflected in the performance of the corresponding artist. Otherwise where is the logic in the acceptance?

4. The raga is of South Indian Classical origin which has been introduced to the North Indian Classical music. For such ragas (e.g. Kirwani and Charukeshi),

   rules regarding Vadi-Samvadi selection are not clear. Our objective analysis seems to be the only way out. Kirwani has been analyzed in [1] itself.

In the present work a two minute vocal recording of raga Rageshree is subjected to a careful statistical analysis. As mentioned earlier, like Bageshree and Shankara, Rageshree is another controversial raga so far as the choice of Vadi-Samvadi selection is concerned where there are two different opinions. Our analysis is broken into three phases: first half, middle half and last half. Under a multinomial model set up holding appreciably in the first two phases, only one opinion is found acceptable.

In the last phase the distribution seems to be quasi multinomial, characterized by an unstable nature of relative occurrence of pitch of all the notes and although the note whose relative occurrence of pitch suddenly shoots is the Vadi swar selected earlier, we take it as an "outlier" demanding a separate treatment. Selection of Vadi-Samvadi notes in a quasi-multinomial set up is still an open research problem. An interesting *musical cocktail* is proposed, however, embedding several ideas like melodic property of notes, note combinations and pitch movements between notes, using some weighted combination of psychological and statistical stability of notes along with watching carefully the sudden shoot of one or more notes whenever there is enough evidence that multinomial model has broken down..

Before we enter the analysis in the next section, it is worthwhile taking a close look at some of the musical features of this raga including the Vadi-Samvadi conflicting opinions.

*Raga: Rageshree*

Thaat (a specific way of grouping ragas): Khambaj
Aroh (ascent): S G M D N **S**
Awaroh (descent): {**S** n D M G} {M R} S
Jati: Aurabh-Sarabh (five distinct notes used in ascent; six
distinct notes used in descent)
Vadi Swar (most important note): G (some say M)
Samvadi Swar (second most important note): N (some say S)
Prakriti (nature): Restful
Pakad (catch): {G M R S}, {*n D*} {*N* S} G M
Nyas swar (Stay notes): G N M S
Time of rendition: 2$^{nd}$ phase of night (9 PM – 12 PM)

**Abbreviations:-**

The letters S, R, G, M, P, D and N stand for Sa (always Sudh), Sudh Re, Sudh Ga, Sudh Ma, Pa(always Sudh), Sudh Dha and Sudh Ni respectively.

The letters r, g, m, d, n represent Komal Re, Komal Ga, Tibra Ma, Komal Dha and Komal Ni respectively.

A note in Normal type indicates that it belongs to middle octave; if in italics it is implied that the note belongs to the octave just lower than the middle octave while a bold type indicates it belongs to the octave just higher than the middle octave.

## 2. Statistical Analysis

Using Solo Explorer 1.0, a software for music transcription cross-checked by our ingenious MATLAB programs, we "exported" the fundamental frequencies against onset times into a text file and detecting the notes adopting a probabilistic approach using *Chebyshev's inequality*. The interested reader is referred to [1] and we omit the details. The artist kept her *Sa* at C as before and the recording done in the Laptop at 44.100 KHz, 16 bit mono, 86 kb/sec mode. As this recording is of longer duration, we could detect more notes. A chief advantage is that the classes do not have to be pooled at least in the first two phases that matter in our present analysis.

Our hypothesis is that the overall distribution (for two minutes) is applicable for smaller parts of the piece as well, that is to say, the three phases. The hypothesis is acceptable for the first two phases only, not the third. Vadi-Samvadi are selected for the first two phases only.

In the following tables, the term "relative frequency" means *ratio of number of occurrences of pitch of a particular note to number of occurrences of pitch of all notes in a specific time period.* Our analysis begins with tables 1-4.

**Table 1: Relative Frequencies for two minutes**

| Note | Relative Frequency |
|---|---|
| Sa | 0.271186 |
| Such Re | 0.135593 |
| Sudh Ga | 0.182203 |
| Sudh Ma | 0.097457 |
| Sudh Dha | 0.046610 |
| Komal Ni | 0.067796 |
| Sudh Ni | 0.199152 |

**Table 2: Relative Frequencies for the first 60 seconds**

| Note | Relative Frequency |
|---|---|
| Sa | 0.281818 |
| Such Re | 0.118181 |
| Sudh Ga | 0.154545 |
| Sudh Ma | 0.090909 |
| Sudh Dha | 0.072727 |
| Komal Ni | 0.072727 |
| Sudh Ni | 0.209090 |

**Table 3: Relative Frequencies for the middle 60 seconds**

| Note | Relative Frequency |
|---|---|
| Sa | 0.295652 |
| Such Re | 0.147826 |
| Sudh Ga | 0.191304 |
| Sudh Ma | 0.060869 |
| Sudh Dha | 0.034782 |
| Komal Ni | 0.043478 |
| Sudh Ni | 0.226086 |

**Table 4: Relative Frequencies for the last 60 seconds**

| Note | Relative Frequency |
|---|---|
| Sa | 0.145454 |
| Such Re | 0.109090 |
| Sudh Ga | 0.290909 |
| Sudh Ma | 0.200000 |
| Sudh Dha | 0.018181 |
| Komal Ni | 0.109090 |
| Sudh Ni | 0.127272 |

We next compute expected frequencies using the formula

Expected frequency of a note in a specific time period = Relative frequency of the note in the specific time period x total number of notes detected in the specific time period

This yields tables 5, 6 and 7.

**Table 5: Expected Frequencies for the first 60 seconds**

| Note | Expected Frequency |
|---|---|
| Sa | 29.830460 |
| Sudh Re | 14.915230 |
| Sudh Ga | 20.042330 |
| Sudh Ma | 10.720270 |
| Sudh Dha | 5.127100 |
| Komal Ni | 7.457560 |
| Sudh Ni | 21.906720 |

**Table 6: Expected Frequencies for the middle 60 seconds**

| Note | Expected Frequency |
|---|---|
| Sa | 31.186390 |
| Sudh Re | 15.593195 |
| Sudh Ga | 20.953345 |
| Sudh Ma | 11.207555 |
| Sudh Dha | 5.360150 |
| Komal Ni | 7.796540 |
| Sudh Ni | 22.902480 |

**Table 7: Expected Frequencies for the last 60 seconds**

| Note | Expected Frequency |
|---|---|
| Sa | 14.915230 |
| Sudh Re | 7.457615 |
| Sudh Ga | 10.021165 |
| Sudh Ma | 5.360135 |
| Sudh Dha | 2.563550 |
| Komal Ni | 3.728780 |
| Sudh Ni | 10.95336 |

We next perform three Chi-Square tests to test the significance of the difference between observed and expected frequencies. The results are shown in tables 8, 9 and 10.

**Table 8; Chi-Square test for the first 60 seconds**

| Note | Observed Frequency(O) | Expected Frequency(E) | O-E | $(O-E)^2$ | $\dfrac{(O-E)^2}{E}$ |
|---|---|---|---|---|---|
| Sa | 31 | 29.830460 | 1.169540 | 1.367823 | 0.045853 |
| Sudh Re | 13 | 14.915230 | -1.915230 | 3.668105 | 0.245930 |
| Sudh Ga | 17 | 20.042330 | -3.042330 | 9.255771 | 0.461811 |
| Sudh Ma | 10 | 10.720270 | -0.720270 | 0.373010 | 0.034794 |
| Sudh Dha | 8 | 5.127100 | 2.872900 | 8.253554 | 1.609789 |
| Komal Ni | 8 | 7.457560 | 0.542440 | 0.294241 | 0.039455 |
| Sudh Ni | 23 | 21.906720 | 1.093280 | 1.195261 | 0.054561 |

Calculated Chi-Square = $\sum ((O-E)^2 / E)$ = 2.492193

Table Chi-Square at 6 degrees of freedom and 5% level of significance = 12.592

**Table 9: Chi-Square test for the middle 60 seconds**

| Note | Observed Frequency(O) | Expected Frequency(E) | O-E | $(O-E)^2$ | $\dfrac{(O-E)^2}{E}$ |
|---|---|---|---|---|---|
| Sa | 34 | 31.186390 | 2.813610 | 7.916401 | 0.253841 |
| Sudh Re | 17 | 15.593195 | 1.406805 | 1.690715 | 0.108426 |
| Sudh Ga | 22 | 20.953345 | 1.046655 | 1.095486 | 0.052282 |
| Sudh Ma | 7 | 11.207555 | -4.207555 | 17.703519 | 1.579605 |
| Sudh Dha | 4 | 5.360150 | -1.36015 | 1.850008 | 0.345141 |
| Komal Ni | 5 | 7.79654 | -2.796540 | 7.820635 | 1.003090 |
| Sudh Ni | 26 | 22.902480 | 3.097520 | 9.594630 | 0.418934 |

Calculated Chi-Square = $\sum ((O-E)^2 / E)$ = 3.761319

Table Chi-Square at 6 degrees of freedom and 5% level of significance = 12.592

**Table 10: Chi-Square test for the last 60 seconds**

| Note | Observed Frequency(O) | Expected Frequency(E) | O-E | $(O-E)^2$ | $\dfrac{(O-E)^2}{E}$ |
|---|---|---|---|---|---|
| Sa | 8 | 14.915230 | -6.915230 | 47.820405 | 3.206145 |
| Sudh Re | 6 | 7.457615 | -1.457615 | 2.124641 | 0.284895 |
| Sudh Ga | 16 | 10.021165 | 5.978835 | 35.746467 | 3.567096 |
| Sudh Ma | 11 | 5.360135 | 5.639865 | 31.808077 | 5.934193 |
| Sudh Dha and Komal Ni | $\left.\begin{array}{l}1\\6\end{array}\right\}7$ | $\left.\begin{array}{l}2.563550\\3.728780\end{array}\right\}6.292330$ | 0.707670 | 0.500796 | 0.079588 |
| Sudh Ni | 7 | 10.953360 | -3.953360 | 15.629055 | 1.426873 |

Calculated Chi-Square = $\sum ((O-E)^2 / E)$ = 14.49879

Table Chi-Square at 5 degrees of freedom and 5% level of significance = 11.070

**Interpretation**

1. Mutinomial model is acceptable at 5% level of significance only for the first two phases as the calculated chi-square values are less than the corresponding tabular value. Hence only the first two phases can be used for Vadi-Samvadi selection. (Independence of notes overall was verified as usual by run test of randomness; see [1]). Multinomial model breaks down in the third phase as the value of calculated chi- Square exceeds the table counterpart. This was only expected if you notice the evident mismatch between tables 1 and 4. At this moment, detecting Vadi-Samvadi notes in the quasi-multinomial set up is an open research problem. See how the relative frequency of even the tonic Sa becomes terribly unstable when multinomial model breaks.

2. We now come to the crucial part of our analysis. Opinions are divided over {G, N} and {M, S} for Vadi Samvadi Selection. Applying condition (i) on tables 1-3, Ma is ruled out from the contest straightaway due to its low relative frequency. This automatically rules out Sa from the contest since Sa's importance is more as a "tonic" (an origin of reference for realizing other notes) and it could at most be a Samvadi Swar, if we could endorse Ma as Vadi (see also the discussion in section 3). Besides, the stability of {G, N} cannot be questioned for the first two phases with G settling to where it should be by the second phase. Both are nyas swars and both have high relative frequencies. Why N has a higher relative frequency than G in the first two phases is a performer dependent phenomenon and it could well be the other way round. A striking feature of the last phase, where multinomial model breaks, is that the Vadi swar G has the highest relative frequency, even higher than Sa. Although we are not analyzing the last phase, we gladly put this highest relative frequency as an "outlier" demanding a separate treatment when we take on quasi-multinomial cases in a subsequent paper.

**3.1 Discussion**

It is important to mention here the strong relationship between our work and that of Krumhansl** [2]. Like us, she has also worked on pitch stability and tonal hierarchy both in Western [3] and North Indian music [2] and, like us, she has also used statistical tools [3]. Like us, she has also admitted the confusion in Vadi-Samvadi selection [2]. The main difference, however, is the way the word "stability" has been interpreted. According to her, *a stay note is a stable note*. Thus Krumhansl's concept of stability is more *psychological* (as stay indicates *rest*) and the stress here is on *note duration* which is one of the four important features in music analysis (the other three are onset, pitch and loudness). Her tonal hierarchy is also created accordingly. Our concept of stability in contrast is *statistical* and we emphasize here that the term (statistically) *stable* in our work means *less fluctuation in relative occurrences of pitch.* All stay notes would be "psychologically stable" but will they all be statistically stable? On the other hand, as mentioned earlier, the Vadi swar of raga Durga of Bilaval thaat, Sudh Ma, whose statistical stability we promise to confirm sometimes later, is not a nyas swar and hence, by Krumhansl's definition, not a psychologically stable note!

For the benefit of the reader, a few lines are quoted below from p. 397 of [2]:

"Like traditional Western music, Indian music is organized around a single tone, the tonic, which is called Sa. …..Sa serves as the starting point of the scale and is the most stable of the scale tones. …Pa is considered the next most stable (except in the few scales that exclude it)".

It is clear from this remark that "stable" in Krumhansl's work [2] (actually a joint work; for convenience, we are referring to it by the corresponding author's name) refers to *psychological stability*, indicating a clear stay on the note, rather than statistical stability which is probabilistic in nature, having to do with *stable relative frequency of pitch occurrence*. Interestingly, although Krumhansl did say "…the prominence of Sa and Pa is common to most ragas whereas the vadi and samvadi are unique to individual ragas", she did not objectively explain why they should be unique. Instead, two confusing definitions are quoted, one from Bhatkhande (vadi is "that note which compared to the other notes in the raga, is sounded most often with clarity", quoted from p. 66 of Jairazbhoy [4]) and another from Danielou (vadi is the "predominant note from which all variations begin and on which they end: it is always accentuated and bears long pauses" quoted from p. 61-62 of [5]). The first definition is quite confusing because first, the note that comes most often, even if it comes with clarity, can easily be Sa, the tonic, which is never the Vadi or it can be some nyas swar which need not be Vadi or Samvadi (like Sudh Re of Yaman) and secondly, if this definition is followed, the vadi becomes exclusively performer dependent whereas it should be *raga dependent endorsed by performance*. The second definition is not acceptable as it forces the vadi to be a nyas swar which is clearly not the case with ragas such as Durga of Bilaval thaat and, more importantly, it suffers from the same drawback as the first definition, namely, with the tonic Sa again causing problems fulfilling the definition easily! Her concluding remark, which is in clear conformity with us, seems to be more correct where she admits "In actuality the employment of the Vadi and samvadi is somewhat ambiguous and their characteristics are still a matter of debate by music theorists"( p. 398 of [2]) and here she again refers to Jairazbhoy [4]).

As a final comment, the concept of statistical stability [1] is important as it can be linked objectively, albeit indirectly, to the role a note plays in expressing the raga while psychological stability[2] addresses the duration of notes and the corresponding tonal hierarchy. The reader should ask himself: "If the psychologically stable Sa were so important in a raga expression, why is there not a single raga where it is the Vadi swar? Is it merely because it is present in every raga?" Well, one of Pa and Ma (Sudh or Tibra) also has to be present in every raga but these notes can be Vadi! The question is worth debating. According to the first author, one explanation is that, *when we say that {M, d, n, P} is an important combination in the raga Kausi Kanada, for example, it is the note combination and how it is to be rendered that are important (this "how" depends on the transitory and non-transitory pitch movements between notes; see more on Varnalankars depicting the "how" part in [1], ref. [6] and Strawn[7]); it matters little where the Sa is taken, whether at C, C# or somewhere else.*.Thus, in the present paper, even though Sa has satisfied condition (i), perhaps even (ii) in the first two phases, it cannot be a Vadi. Even the selection of Sa as a Samvadi swar is done when no other note can contest to be Samvadi as also to maintain the Vadi-Samvadi distance (ideally they should be separated by an interval of a fifth). The most important note, musically speaking, must be *the one which plays the most significant role in raga expression*. We will take this as an informal definition (though subjective because of the clause "which plays the most significant role in raga expression"). [1] briefs the role of Vadi in deciding whether the raga is *purvanga pradhan* (first half more important) or *uttaranga pradhan* (second half more important) and in deciding broadly the time of raga rendition. Statistically speaking, it should be the note, of course different from the tonic Sa for reasons stated earlier, satisfying conditions (i) and (ii) at least under a multinomial set up. It should be understood that this statistical tool is only an objective indirect strategy that challenges a subjective as well as ambiguous issue and seeks to explore some truths, and is itself by no means a definition. In the second part of this section, we provide Krumhansl's method of analysis of Rageshree taking note duration into consideration.

## 3.2 Krumhansl's method of analysis by note-duration: psychological stability and tonal hierarchy

Caution: This method only detects the stay notes or nyas swars but not the Vadi-Samvadi swars. For detecting Vadi-Samvadi swars, statistical stability is a better concept. However, in ragas where the Vadi-Samvadi are known to be nyas swars as in Rageshree, *Krumhansl's technique does help in reducing the search space*. We are seriously thinking of using a weighted combination of both concepts to create a concept of *universal stability* (see sec 4), given that ragas where Vadi Samvadi are themselves not nyas swars are few (e.g. Durga of Bilaval thaat).

Table 11:Summary table of note-duration

| Note | Mean duration (sec) | Standard deviation of duration (sec) | Rank (based on mean duration) |
|---|---|---|---|
| Sa | 0.212173 | 0.365006 | 02 |
| Sudh Re | 0.181304 | 0.307078 | 04 |
| Sudh Ga | 0.197083 | 0.302233 | 03 |
| Sudh Ma | 0.042272 | 0.024839 | 05 |
| Sudh Dha | 0.023000 | 0.010049 | 06 |
| Sudh Ni | 0.228275 | 0.403301 | 01 |
| Komal Ni | 0.021875 | 0.012806 | 07 |

Remark: Ranks are given on the basis of mean duration. Higher rank (i.e. lower numerical value; rank 01 is higher than rank 02) is given for higher mean duration. If mean durations are equal (say upto six places of decimals), the note with a lower standard deviation will be given a higher rank.

Interpretation:

The nyas swars experimentally detected using Krumhansl's theory are Sudh Ni, Sa, Sudh Ga and Sudh Re. Theoretically they should be Sudh Ga, Sudh Ni, Ma and Sa. Thus Ma does not have a good rank to claim psychological stability. Since the Vadi in Rageshree has to be a nyas swar, Ma loses the contest for being Vadi from psychological consideration of note stability. As we have already ruled it out from statistical considerations earlier, the techniques of Krumhansl and Chakraborty are both confirming the rejection of Ma as Vadi. For reasons stated earlier, Sa is also rejected for being Samvadi. This means the Vadi-Samvadi can only be (Sudh Ga, Sudh Ni). And we select Ga as Vadi and Ni as Samvadi from statistical considerations as stated earlier.

We are, however, not ruling out {M, S} for all situations, this being a case study. We only assert that, with the present performance, *the artist can only defend that musical school which supports {G, N} as Vadi-Samvadi.* This concludes our discussion.

## 4. Conclusion and future work

The present performance supports {G, N} as Vadi-Samvadi notes for raga Rageshree, under a multinomial set up. Regarding how to detect the Vadi-Samvadi in a quasi-multinomial set up, the most general situation, we are seriously thinking about combining the concept of melodic property of notes, the note combinations detected in the performance involving a specific note, the *Varnalankars*, and some weighted combination of both concepts of stability, psychological and statistical.

"...we must develop a concept of universal stability by combining the concepts of psychological stability and statistical stability. The best way is to do two different rankings first and then assign an average rank which will NOT be the arithmetic mean but weighted mean of the two ranks (one rank emerging from your method;the other from my method).The research problem then is to decide what the weights should be.

Let me make things clear with an example. If you study the attached paper (this very paper), you will find the note Sudh Ga has been selected as Vadi using the concept of statistical stability and hence it has rank 01 by my method. But it has been assigned rank 03 by your method of analysis (psychological stability)! It is therefore important that a rank of universal stability be assigned to this note. If you take arithmetic mean of the two ranks, then the rank 02 should be assigned to it. If you go for weighted mean (which is more rational; AM is only a special case thereof) then this note should be assigned the rank $(01*w_1+03*w_2)/(w_1+w_2)$. Here $w_1$ is the weight or importance corresponding to a rank that signifies statistical stability. Similarly $w_2$ is the weight or importance corresponding to a rank of psychological stability .Our problem is to decide the values of $w_1$ and $w_2$. Here you have to bear in mind that statistical stability is more helpful in deciding the Vadi-Samvadi and it is a count-based analysis. On the other hand, psychological stability is more helpful in detecting the nyas swars or stay notes and it is a duration based analysis. If our research is meant to detect Vadi-Samvadi, we must keep $w_1>w_2$ while if the research is aimed at detecting the nyas swars, then we must keep $w_2>w_1$....." [08]

Additionally, we are also not taking the sudden "shoot" of the Vadi in the third phase lightly. Outliers are never thrown away in statistics. Rather statisticians try to find out how they are caused and what they signify. In fact, there is a separate branch of statistics to deal with them [09]. When multinomial model breaks in other situations, we shall closely watch the notes that "shoot". As in [1], we again feel it important to mention here that a raga is not merely a melodic structure but in addition one with a set of rules characterizing a certain mood endorsed through performance. The rules are created not to restrict the performer but to characterize the underlying mood and it is possible to break the rules and yet increase the melodic properties of notes (for example, by including some additional notes in the raga which destroys its purity or by killing "classicism" by "romanticism"-more about it some other day) and hence a conjecture such as "Vadi is that note which is melodically the heaviest" will not do. It would be wonderful if this conjecture worked, as melodic property of notes can be objectively measured, perhaps best by the RUBATO software which assigns weights to notes using all the four features (onset, pitch, duration and loudness) (see [10] for a use of this software in North Indian music). The morale is that a *musical cocktail* will be our response to the complex and unpredictable quasi-multinomial situation. We close this paper with an inter-onset interval graph (fig.1), often used by musicologists to study rhythm, and a transition count (tables 12 and 13) depicting the Varnalankars [1][6][7].

Fig. 1: Inter onset interval graph for rageshree:

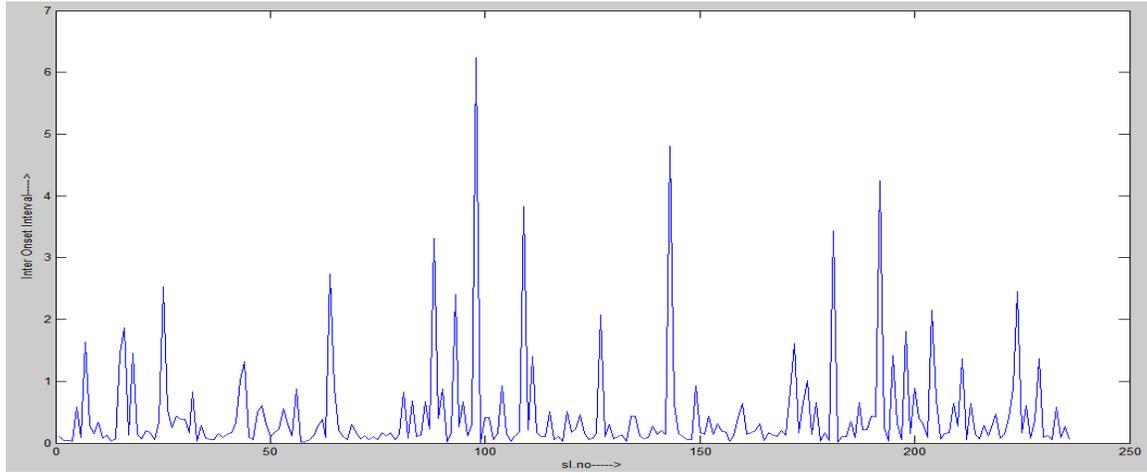

X-axis gives serial number of successive note pair. Y-axis gives inter-onset interval times, i.e. time between two successive onsets. Notes are in rhythm if the inter-onset interval times are equal.

Transition count for Rageshree:-

Table 12: Table of rising , falling, mixed and no transition

| Rising transitions | | | Falling Transitions | | | Mixed transitions | No transition |
|---|---|---|---|---|---|---|---|
| 60 | | | 62 | | | 05 | 28 |
| Convex | Concave | Linear | Convex | Concave | Linear | | |
| 29 | 17 | 14 | 15 | 16 | 31 | | |

Table 13: Table of hats and valleys

| Hats shaped as "^" | | | Valleys shaped as "v" | | |
|---|---|---|---|---|---|
| 103 | | | 113 | | |
| Positively skewed | Negatively skewed | Symmetric | Positively skewed | Negatively skewed | Symmetric |
| 22 | 42 | 39 | 47 | 16 | 50 |
| Low | Moderate | High | Shallow | Moderate | Deep |
| 43 | 50 | 10 | 64 | 41 | 8 |

Interpretation: Almost equal rising and falling transitions implies the raga cannot be classified as one of *arohi* or *awarohi varna* from the performance. The meaning of hats and valleys is under investigation.

**Remark**: There was an error in version 1 (v1) of the archive paper[1]. It was wrongly stated that the nyas swars of Yaman are Sudh Re, Sa, sudh Ni and Pa wheras they are actually Sudh Re, Sudh Ga, Sudh Ni and Pa. Hence sudh Ga, the Vadi in Yaman, is very much a nyas swar. *The raga where the Vadi is itself not a nyas swar is raga Durga of Bilaval that, as mentioned earlier, where Vadi: Sudh Ma; Samvadi:Sa; nyas swars: Sudh Re, Pa and Sudh Dha. We have uploaded a replacement paper (v2).*

[Concluded]

**Acknowledgement:** We wholeheartedly thank Dr. C. Krumhansl (Department of Psychology, Cornell Univ., USA) for letting us have a copy of her paper [2] without which this paper could not be prepared. Interestingly, her ideas of pitch stability and tonal hierarchy were motivated from North Indian ragas! [11]. It is also a pleasure to thank the artist for the vocal rendition of Rageshree.

## Appendix

**Multinomial distribution:** Consider n independent trials being performed. In each trial the result can be any one of k mutually exclusive and exhaustive outcomes e1, e2…ek, with respective probabilities p1, p2…pk. These probabilities are assumed fixed from trial to trial. Under this set up, the probability that out of n trials performed, e1 occurs x1 times, e2 occurs x2 times ….ek occurs xk times is given by the well known multinomial law

$\{n!/(x1! \, x2!...xk!)\} p1^{x1} p2^{x2} \ldots pk^{xk}$ where each xi is a whole number in the range 0 to n subject to the obvious restriction on the xi's, namely, x1+x2+…..+xk=n. For this distribution, E(xi)=n(pi), Var(xi)=n(pi)(1-pi), i=1, 2…….k. and Cov (xi, xj) = -n(pi)(pj). One can now calculate the correlation coefficient between xi and xj as Cov(xi, xj)/√[Var(xi)*Var(xj)].

Varying probabilities lead to a *quasi multinomial distribution*.